\documentclass[aps,twocolumn,preprintnumbers,amsmath,amssymb,superscriptaddress,floatfix,nofootinbib]{revtex4}

\usepackage{lipsum}
\usepackage{graphicx}
\usepackage{epsfig}
\usepackage{epstopdf}
\usepackage{hyperref}
\usepackage{amsmath}
\usepackage{amsfonts}
\usepackage{amssymb}

\usepackage[normalem]{ulem}  
\usepackage[dvipdfmx]{color} 
\usepackage{float}

\begin{document}

\title{Role of a triangle singularity in the $\pi N(1535)$ contribution to $\gamma p \to p \pi^0 \eta$}

\author{V.~R.~Debastiani}
\email{vinicius.rodrigues@ific.uv.es}
\affiliation{Departamento de
F\'{\i}sica Te\'orica and IFIC, Centro Mixto Universidad de
Valencia-CSIC Institutos de Investigaci\'on de Paterna, Aptdo.
22085, 46071 Valencia, Spain}

\author{S.~Sakai}
\email{shuntaro.sakai@ific.uv.es}
\affiliation{Departamento de
F\'{\i}sica Te\'orica and IFIC, Centro Mixto Universidad de
Valencia-CSIC Institutos de Investigaci\'on de Paterna, Aptdo.
22085, 46071 Valencia, Spain}

\author{E.~Oset}
\email{oset@ific.uv.es}
\affiliation{Departamento de
F\'{\i}sica Te\'orica and IFIC, Centro Mixto Universidad de
Valencia-CSIC Institutos de Investigaci\'on de Paterna, Aptdo.
22085, 46071 Valencia, Spain}

\date{\today}

\begin{abstract}
We have studied the $\gamma p \to p \pi^0 \eta$ reaction paying attention to the two main mechanisms at low energies, the $\gamma p \to \Delta(1700) \to \eta \Delta(1232)$ and the $\gamma p \to \Delta(1700) \to \pi  N(1535)$. Both of them are driven by the photoexcitation of the $\Delta(1700)$ and the second one involves a mechanism that leads to a triangle singularity. We are able to evaluate quantitatively the cross section for this process and show that it agrees with the experimental determination. Yet, there are some differences with the standard partial wave analysis which does not include explicitly the triangle singularity. The exercise also shows the convenience to explore possible triangle singularities in other reactions and how a standard partial analysis can be extended to accommodate them.
\end{abstract}

\maketitle

\section{Introduction}

The $\gamma p \to p \pi^0 \eta$ reaction was measured first in Ref. \cite{Nakabayashi:2006ut} up to energies of the photon of $E_{\gamma} = 1150$ MeV. Early theoretical determinations of the threshold behaviour, with large uncertainties were done in Ref. \cite{Jido:2001nt}. Some accurate predictions in the range up to $E_{\gamma} = 1700$ MeV, were done in Ref. \cite{Doring:2005bx} prior to the measurements done at GRAAL \cite{Ajaka:2008zz}, CB-ELSA \cite{Horn:2007pp,Horn:2008qv} and MAMI \cite{Kashevarov:2009ww}. The basic idea of Ref. \cite{Doring:2005bx} was that the process is dominated by the photoproduction of the $\Delta(1700)(3/2^-)$, which later decays into $\eta \Delta(1232)$ followed by $\Delta(1232) \to \pi^0 p$. The dominance of this resonance at low energies was also established experimentally \cite{Ajaka:2008zz,Horn:2008qv,Kashevarov:2009ww}. In Ref. \cite{Kashevarov:2009ww} it is quoted ``it is possible to get a reasonable agreement with the data by taking into account only the $D_{33}(1700)$ resonance''. Further support for this idea comes from the correlation of many reactions based upon the dominance of the $\Delta(1700)$. Indeed, in Ref. \cite{Doring:2006pt} the $\pi^- p \to K^0 \pi^0 \Lambda$, $\pi^+ p \to K^+ \pi^+ \Lambda,\; K^+ \bar K^0 p, \; K^+ \pi^+ \Sigma^0, \; K^+ \pi^0 \Sigma^+, \; \eta \pi^+ p$ reactions were described successfully based upon the mechanism of $\Delta(1700)$ excitation with subsequent decays into $K \Sigma^*(1385)$ or $\eta \Delta(1232)$. The $p \pi^0$, $p \eta$ and $\eta \pi^0$ mass distributions measured in Ref. \cite{Ajaka:2008zz} also give support to this idea, which is further reinforced by the agreement shown in Ref. \cite{Doring:2010fw} for the polarization observables $I^S$ and $I^C$, $I^\theta$ measured in Refs. \cite{Gutz:2008zz,Gutz:2009zh}.

A high statistics measurement of different observables is done in Ref. \cite{Gutz:2014wit}. In this work a separation of the cross section is made in three main channels, $\eta \Delta$, $\pi N(1535)$ and $a_0(980) p$, and up to $E_{\gamma}$ around 1500 MeV the first two channels saturate the cross section. The $\eta \Delta$ channel is dominant, but the $\pi N(1535)$ is also sizable in this region. The purpose of the present work is to find a theoretical description of these two channels.

The $\eta \Delta$ channel finds a natural interpretation in the dominance of the $\Delta(1700)$ excitation and provides support for the dynamical generation of this resonance from the interaction of the octet of pseudoscalar mesons with the decuplet of baryons \cite{Kolomeitsev:2003kt,Sarkar:2004jh}. Indeed, as shown in Ref. \cite{Sarkar:2004jh}, the $\Delta(1700)$ is generated from the coupled channels $\Delta \pi$, $\Sigma^* K$ and $\Delta \eta$, and the scattering matrix leads to a sizable coupling of that resonance to $\Delta \eta$. Hence, the main channel assumed in Ref. \cite{Doring:2005bx} is photoproduction of the $\Delta(1700)$ followed by the decay of the $\Delta(1700)$ into $\Delta \eta$ and posterior $\Delta \to \pi N$ decay. In this mechanism there is no direct room for the $\pi  N(1535)$ channel, although some terms, with final state interaction of $\pi \eta$, partly incorporated this channel in Ref. \cite{Doring:2005bx}. In the present work we are going to show that the relatively large weight of the $\pi N(1535)$ channel is tied to a triangular singularity for the process $\gamma p \to \Delta(1700) \to \eta \Delta^+$ followed by $\Delta^+ \to \pi^0 p$ and posterior fusion of the $p \eta$ to produce the $N(1535)$.

Triangle singularities were first discussed by Landau \cite{Landau:1959fi}, but it is now, with the large amount of experimental information gathered on particle reactions and resonances, that the relevance of the idea in hadron physics has become apparent. Some examples of triangle singularities are shown in Ref. \cite{Liu:2015taa}. In essence this consists of a particle $A$ decaying into $1+2$, particle 2 decaying into 3 and $B$ (external), and particles $1+3$ fusing to give another external particle $C$. The singularity appears when the former process occurs at a classical level, which is stated in terms of the Coleman-Norton theorem \cite{Coleman:1965xm}. An easy and practical way to show when a singularity appears is given in Ref. \cite{Bayar:2016ftu}, where a different approach to the standard one is followed.

Other examples of triangle singularities can be seen in Refs. \cite{Wang:2013hga,Achasov:2015uua,Lorenz:2015pba,Szczepaniak:2015eza,Szczepaniak:2015hya}. More closely related to the present problem is the case of the $\eta(1405) \to \pi a_0(980), \; \pi f_0(980)$ \cite{Wu:2011yx,Aceti:2012dj,Wu:2012pg}, where in particular, the latter channel violating isospin is enhanced due to a triangle singularity. Another recent example can be seen in the ``$a_1(1420)$'' peak, originally advocated as a new resonance by the COMPASS collaboration, which hinted in Ref. \cite{Liu:2015taa} and shown explicitly in Refs. \cite{Ketzer:2015tqa,Aceti:2016yeb}, comes naturally from the $\pi f_0(980)$ decay of the $a_1(1260)$, via a triangular mechanism that develops a singularity when the $a_1(1260)$ decays into $K^* \bar K$, the $K^* \to \pi K$ and the $K \bar K$ merge to produce the $f_0(980)$. A case similar to this is the recent reanalysis of the $f_1(1420)$, which is shown in Ref. \cite{Debastiani:2016xgg} to correspond to two mechanisms: the decay of the $f_1(1285)$ into $\pi a_0(980)$ via a triangular singularity, $f_1(1285) \to K^* \bar K$, $K^* \to \pi K$, $K \bar K \to a_0(980)$; and decay into $K^* \bar K$, that shows as a pronounced peak above the $K^* \bar K$ threshold. Adding to this list of reinterpretation of some accepted resonances is the case of the $f_2(1810)$, also explained in Ref. \cite{Xie:2016lvs} as the production of the $f_2(1650)$ followed by the decay into $K^* \bar K^*$, $K^* \to \pi K$, $K \bar K^* \to a_1(1260)$.

In some cases the singularity helps to explain enhancements in cross sections not attributed to any resonance. This is the case of the $\gamma p \to K \Lambda(1405)$ reaction, where a triangular singularity stemming from the production of a $N^*$ resonance at 1930 MeV, with $N^* \to K^* \Sigma$, $K^* \to K \pi$ and $\pi \Sigma$ merging to give the $\Lambda(1405)$, produces a peak in the cross section around $\sqrt{s} = 2120$ MeV \cite{Wang:2016dtb}, that solves a problem in the interpretation of the data \cite{Moriya:2013hwg}.

Recent interest in triangle singularities was stirred by the suggestion in Refs. \cite{Guo:2015umn,Guo:2016bkl} that the peak seen in the LHCb collaboration attributed to a pentaquark in Refs. \cite{Aaij:2015tga,Aaij:2015fea} should be due to a triangle singularity stemming from $\Lambda_b \to \Lambda(1890) \, \chi_{c1}$, $\Lambda(1890) \to K^- \, p$, $\chi_{c1} \, p \to J/\psi \, p$. However, the $\chi_{c1} \, p$ system is at threshold for the energy of the peak at 4450 MeV, and if this peak has quantum numbers $3/2^-$ or $5/2^-$ as suggested by the experiment, the $\chi_{c1} \, p$ system must be in $P$- or $D$-wave, which at threshold kills the $\chi_{c1} \, p \to J/\psi \, p$ amplitude. This observation was made in Ref. \cite{Bayar:2016ftu} where it was concluded that this mechanism could not be the explanation of the experimental peak if these quantum numbers are confirmed.

In the present work we will show another case of a triangle singularity via $\gamma p \to \Delta(1700) \to \eta \Delta \to \eta \pi^0 p$, with $\eta \, p$ merging into the $N(1535)$, which gives rise to a $\pi N(1535)$ production cross section similar in strength and shape to the experimental one. We will also show that the energy dependence of this cross section is quite different to a standard one proceeding thorough $\gamma p \to \pi N(1535)$ directly, and it is tied to the structure of the triangle singularity.

\section{Formalism}

\subsection{The tree level $\gamma p \to \Delta(1700) \to \Delta \eta$}

\begin{figure}[H]
 \centering
 \includegraphics[width=0.5\textwidth]{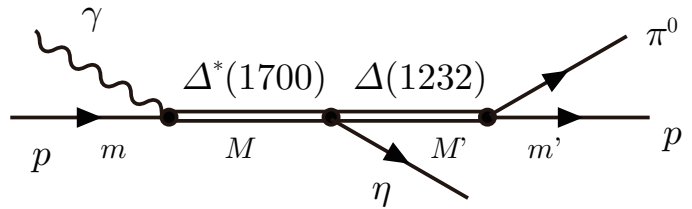}
 \caption{Mechanism for $\gamma p \to \Delta(1700) \to \eta \Delta(1232) \to \eta \pi^0 p$ driven by $\Delta(1700)(3/2^-)$ photoproduction.}
 \label{fig_1}
\end{figure}

In Fig.~\ref{fig_1} we depict the mechanism for direct production of the $\Delta(1700)$ followed by the decay into $\Delta(1232) \eta$ and $\Delta(1232) \to \pi^0 p$.

The first ingredient needed in the evaluation is the $\Delta(1700) \gamma p$ coupling. This proceeds in $S$-wave and can be taken into account with the amplitude
\begin{align}\label{eq_7_1}
 -it_{\Delta^*, \gamma p}=-ig_{\Delta^*, \gamma p} \, \vec{S}\cdot\vec{\epsilon}
\end{align}
where $\vec{\epsilon}$ is the polarization of the photon in the Coulomb representation ($\epsilon^0=0$) and $\vec{S}$ the spin transition operator from $3/2$ to $1/2$.
The width of $\Delta^*$ into this channel is given by
\begin{align}
 \Gamma_{\Delta^*,\gamma
 p} = \, &\frac{1}{2\pi}\frac{M_N}{M_{\Delta^*}}p_\gamma\overline{\sum}\sum|t_{\Delta^*,\gamma
 p}|^2
 \label{eq_8_1},
\end{align}
where

\begin{widetext}\vspace{-0.2in}
\begin{align}
 \overline{\sum}\sum|t_{\Delta^*,\gamma p}|^2
 =\, &|g_{\Delta^*,\gamma p}|^2\, \frac{1}{4}\sum_M \sum_m \sum_{\gamma\ {\rm pol}} \left<m\left|\,\vec{S}\cdot\vec{\epsilon}\,\right|M\right>
 \left<M\left|\,\vec{S}^\dagger\cdot\vec{\epsilon}\,\right|m\right>\notag\\
 = \, &|g_{\Delta^*,\gamma p}|^2 \,\frac{1}{4} \sum_m \sum_{\gamma\ {\rm pol}} \left<m\left|\frac{2}{3}\,\delta_{ij}-\frac{i}{3}\epsilon_{ijk}\sigma_k\right|m\right>
 \epsilon_i\epsilon_j\notag\\
 = \, &\frac{1}{3}\,|g_{\Delta^*,\gamma p}|^2 \sum_{\gamma\ {\rm pol}} \vec{\epsilon}\cdot\vec{\epsilon}
 = \frac{2}{3}\,|g_{\Delta^*,\gamma p}|^2.
 \label{eq_8_3}
\end{align}
\end{widetext}

Experimentally, the branching fraction is $0.22 - 0.60$\% from a Breit-Wigner width $\Gamma_{\Delta^*}=200-400$ MeV and mass $M_{\Delta^*}=1670-1750$ MeV \cite{pdg}. We shall play with the uncertainties for a more accurate fit to the $\gamma p \to p \pi^0 \eta$ data. By taking the central value of the branching fraction, 0.41\%, $\Gamma_{\Delta^*}=300$ MeV and $M_{\Delta^*}=1700$ MeV, we obtain from Eq.~(\ref{eq_8_1})
\begin{align}
 g_{\Delta^*,\gamma p}=0.188.\label{eq_8_2}
\end{align}

The PDG has also data for the helicity amplitudes. It is easy to construct the helicity amplitudes from the coupling of Eq.~(\ref{eq_7_1}), following the steps of Ref. \cite{Sun:2010bk}, and show that both, helicity $1/2$ and $3/2$, are compatible with the structure of Eq.~(\ref{eq_7_1}) and the coupling of Eq.~(\ref{eq_8_2}). On the other hand, the coupling of $\Delta^*$ to the channel $\eta \Delta$ is one of the outputs of the chiral unitary approach of Ref. \cite{Sarkar:2004jh} where we find
\begin{align}
 g_{\Delta^*,\eta\Delta}=1.7-i\,1.4 \,,
\end{align}
and the amplitude $-it_{\Delta^*,\eta\Delta}$ is just $-ig_{\Delta^*,\eta\Delta}$ since the process proceeds via $S$-wave and has no spin dependence.

The $\Delta(1232)$ decaying to $\pi^0 N$ has a standard coupling as
\begin{align}
 -it_{\Delta,\pi N}=\frac{f_{\pi N\Delta}}{m_{\pi}} \,\vec{S}\cdot\vec{k}\,\;
 \mathcal{C}(1,1/2,3/2 \,;\,i_\pi,i_N,i_\Delta),
\end{align}
with $\mathcal{C}(1,1/2,3/2\,;\,i_\pi,i_N,i_\Delta)$ the Clebsch-Gordan coefficient, $\sqrt{2/3}$ for $\pi^0 p$, and $\vec{k}$ the pion momentum. From the $\Delta$ width, we find
\begin{align}
 \frac{f_{\pi N\Delta}^2}{4\pi}=0.36\,; \,\ f_{\pi N\Delta}=2.13.
\end{align}

The amplitude of Fig.~\ref{fig_1} can now be constructed with the former ingredients and we have
\begin{widetext}
\begin{align} \label{eq_10_1}
 -it_{\gamma p,\eta\pi^0p}
 =& \sum_M\sum_{M'} \frac{f_{\pi N\Delta}}{m_\pi}
 \left<m'\left| \vec{S}\cdot\vec{k} \right|M'\right> \sqrt{\frac{2}{3}}\,
 \frac{i}{M_{\rm inv}(\pi^0 p) -M_{\Delta} +i\,\Gamma_{\Delta}/2}\notag\\
 &\times(-i\,)\, g_{\Delta^*,\eta\Delta} \,\delta_{MM'}\,
 \frac{i}{\sqrt{s} -M_{\Delta^*} +i\,\Gamma_{\Delta^*}/2}\, (-i\,)\, g_{\Delta^*,\gamma p}
 \left<M\left| \,\vec{S}^\dagger\cdot\vec{\epsilon}\, \right|m\right>\notag\\
 =& \; g_{\Delta^*,\gamma p} \; g_{\Delta^*,\eta\Delta} \,\frac{f_{\pi N\Delta}}{m_\pi}\,
 \sqrt{\frac{2}{3}} \,\frac{1}{M_{\rm inv}(\pi^0p) -M_\Delta +i\,\Gamma_\Delta/2} \notag\\
 & \times \frac{1}{\sqrt{s} -M_{\Delta^*} +i\,\Gamma_{\Delta^*}/2}
 \sum_{M}\left<m'\left| \,\vec{S}\cdot\vec{k}\, \right|M\right>
 \left<M\left| \,\vec{S}^\dagger\cdot\vec{\epsilon}\, \right|m\right> ,
\end{align}
\end{widetext}
where $M_{\rm inv}(\pi^0p)$ is the invariant mass of the $\pi^0p$ system and
$s$ the ordinary Mandelstam variable for the center-of-mass (CM)
energy of the $\gamma p$ initial system.

\subsection{The triangle singularity in $\gamma p \to \pi^0 N(1535)$}

\begin{figure}[H]
 \centering
 \includegraphics[width=0.5\textwidth]{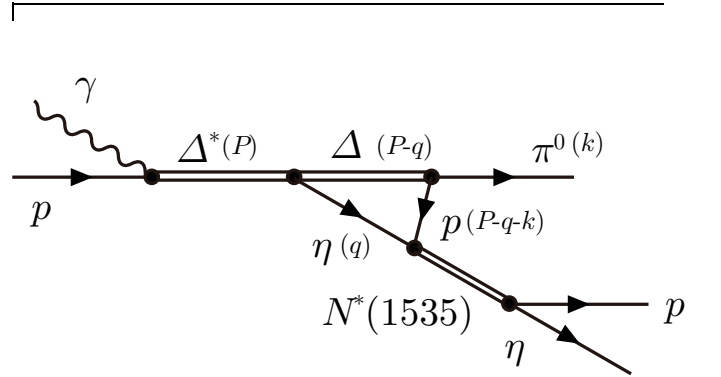}
 \caption{Triangle diagram leading to the production of $\pi N(1535)$ ($\eta p$). In parenthesis, the momenta of the particles.}
 \label{fig_2}
\end{figure}

The mechanism that we shall study is depicted in Fig.~\ref{fig_2}. The $\Delta^*$ decays into $\Delta\eta$, the $\Delta$ decays into $\pi^0 p$ and the $\eta p$ merge to produce the $N(1535)$ that subsequently decays into $\eta p$. It is easy to see, by taking Eq.~(18) of Ref. \cite{Bayar:2016ftu} that the diagram of Fig.~\ref{fig_2} develops a singularity around $\sqrt{s}=1782$ MeV, which corresponds to $E_\gamma$ in the laboratory frame at $1220$ MeV. One can see in Ref. \cite{Gutz:2014wit} that there is some kind of broad structure around $E_\gamma=1200$ MeV for the $\pi N(1535)$ part of the cross section in the analysis done there.

 The amplitude for the mechanism of Fig.~\ref{fig_2} is given by
\begin{widetext}
\begin{align}
 -it= &-it_{\eta p,\eta p}\, \frac{f_{\pi  N\Delta}}{m_\pi}\, \sqrt{\frac{2}{3}}\, \vec{S}\cdot\vec{k}\,
  (-i\,)\, g_{\Delta^*,\eta\Delta} \,(-i\,)\, g_{\Delta^*,\gamma p} \;\vec{S}^\dagger\cdot\vec{\epsilon}\,
  \frac{i}{\sqrt{s} -M_{\Delta^*} +\,i\Gamma_{\Delta^*}/2}\notag\\
 &\times \int\frac{d^4q}{(2\pi)^4}\, 2M_\Delta\frac{i}{(P-q)^2 -M_\Delta^2 +i\,\epsilon} \,2M_N\,
 \frac{i}{(P-q-k) -M_N^2 +i\epsilon}\, \frac{i}{q^2 -m_\eta^2 +i\epsilon}\notag\\
 =&\;t_{\eta p,\eta p} \,g_{\Delta^*,\eta \Delta} \,g_{\Delta^*,\gamma p} \,\frac{f_{\pi N\Delta}}{m_\pi}
 \,\sqrt{\frac{2}{3}}\, \vec{S}\cdot\vec{k} \; \vec{S}^\dagger\cdot\vec{\epsilon}\;
 2M_N \, 2M_\Delta\, \frac{1}{\sqrt{s} -M_{\Delta^*} +i\,\Gamma_{\Delta^*}/2}\,t_T, \label{eq_11_1}
\end{align}
\end{widetext}
which defines $t_T$, the triangle amplitude, as $i\int d^4q$ of the product of the three propagators, $\Delta$, $\eta$, $p$. The Mandl-Shaw normalization for fermion fields \cite{MandlShaw:2010}, responsible for the factors $2M_N$, $2M_\Delta$ in Eq.~(\ref{eq_11_1}), is used. The $q^0$ integration in Eq.~(\ref{eq_11_1}) is done analytically and then the $t_T$ amplitude is written as \cite{Aceti:2015zva,Bayar:2016ftu}
\begin{widetext}
\begin{align}
 t_T=\int\frac{d^3q}{(2\pi)^3}&\,\frac{1}{8\omega(q)\omega'(q)\omega^*(q)}
 \,\frac{1}{k^0 -\omega'(q) -\omega^*(q) +i\epsilon}
 \,\frac{1}{P^0-\omega^*(q)-\omega(q)+i\epsilon}\notag\\
 &\,\frac{2P^0\omega(q) +2k^0\omega'(q) -2[\omega(q)+\omega'(q)][\omega(q) +\omega'(q) +\omega^*(q)]}{(P^0 -\omega(q) -\omega'(q) -k^0 +i\epsilon)(P^0 +\omega(q) +\omega'(q) -k^0 -i\epsilon)} \, , \label{eq_12_1}
\end{align}
where $\omega(q)=\sqrt{{m_\eta}^2+\vec{q\,}^2}\,$,
$\omega'(q)=\sqrt{{M_N}^2+(\vec{q}+\vec{k})^2}\,$,
$\omega^*(q)=\sqrt{{M_\Delta}^2+\vec{q\,}^2}\,$.
\end{widetext}

To account for the width of the $\Delta$ in the loop function we replace $\omega^*(q) \to \omega^*(q) -i\,\Gamma_\Delta/2$, where we use an energy-dependent width
\begin{align}\label{eq:EdepWidth}
\Gamma_\Delta (M_{\rm inv}(q)) = \frac{M_\Delta}{M_{\rm inv}(q)}\frac{{p_\pi}^3(M_{\rm inv}(q))}{{p_\pi}^3|_{\rm on}}\Gamma_\Delta|_{\rm on},
\end{align}
where $M_{\rm inv}(q)$ is the $\Delta(1232)$ invariant mass inside the triangular loop, calculated from the second denominator of Eq.~(\ref{eq_12_1})
\begin{align}
M_{\rm inv}(q) = \sqrt{(P^0)^2 + {m_\eta}^2 -2P^0\,\omega(q)},
\end{align}
and $\Gamma_\Delta|_{\rm on} = 117$ MeV, while $p_\pi(M_{\rm inv}(q))$ is the pion momentum in the $\Delta(1232)$ rest frame
\begin{align}
p_\pi (M_{\rm inv}(q)) &= \frac{\lambda^{1/2}(M_{\rm inv}(q)^2,\, {M_N}^2,\, {m_\pi}^2)}{2M_{\rm inv}(q)},\\
p_\pi|_{\rm on} &= p_\pi (M_{\rm inv}(q)=M_\Delta),
\end{align}
where $\lambda$ is the K\"{a}llen function, and we set $\Gamma_\Delta$ to zero if $M_{\rm inv}(q) < M_N + m_\pi$.

The $\eta p \to \eta p$ amplitude $t_{\eta p,\eta p}$ is driven by the $N(1535)$, which also shows up as a dynamically generated resonance in Ref. \cite{Inoue:2001ip}. The integral in Eq.~(\ref{eq_12_1}) is convergent. Yet, one must take into account that the chiral unitary approach of Ref. \cite{Sarkar:2004jh} can be formally obtained using a Quantum Mechanical
formulation with a potential of the type $V(\vec{q}\, , \vec{q}\,') =
V\theta(q_{\rm max} -|\vec{q}\,|) \theta(q_{\rm max} -|\vec{q}\,'\,|)$
\cite{Gamermann:2009uq} and this leads to a $T$-matrix where the two
$\theta$ functions are also factorized, leading to a $q_{\rm max}$ in the $d^3q$ integration of Eq.~(\ref{eq_12_1}). A value of $q_{\rm max}$ suited of the $\Delta^* \to \Delta\eta$ as well as for the $t_{\eta p,\eta p}$ that we take from the work of Ref. \cite{Inoue:2001ip} done along similar lines, is $q_{\rm max}=800$ MeV in the $N(1535)$ rest frame, that we shall use in our study.
We can see that the amplitude of the triangle diagram, Eq.~(\ref{eq_11_1}) and the tree level $\gamma p \to \Delta^* \to \Delta \eta$ of Eq.~(\ref{eq_10_1}) have exactly the same spin structure $\vec{S}\cdot\vec{k} \; \vec{S}^\dagger\cdot\vec{\epsilon}$, and we expect some kind of interference, although the amplitudes are complex and one has to see explicitly how the interference occurs.

The cross section for $\gamma p \to \eta \pi^0 p$ is given by the standard $a+b \to 1+2+3$ formalism, as
\begin{align}
 \sigma=&\frac{(2M_N)^2}{4p_\gamma\sqrt{s}} \int\frac{d^3p_1}{(2\pi)^3} \frac{1}{2E_1} \frac{1}{16\pi^2} \nonumber \\
 &\times \int d\tilde{\Omega}_2 \, \tilde{p}_2 \frac{1}{M_{\rm inv}(23)} \overline{\sum} \sum|T|^2, \label{eq_13_1}
\end{align}
where $\tilde{p}_2$ is the momentum of particle $2$ in the rest frame of
$2+3$ and $\tilde{\Omega}_2$ its solid angle in that frame.
Proceeding like in Eq.~(\ref{eq_8_3}), summing over transverse photons
\begin{align}
 \sum_{\gamma\ {\rm pol}}\epsilon_i \, \epsilon_j
 =\delta_{ij} -\frac{p_{\gamma i} \, p_{\gamma j}}{\vec{p}_\gamma\,^2},\label{eq_13_2}
\end{align}
we find that
\begin{align}
 \overline{\sum}\sum|\vec{S}\cdot\vec{k} \; \vec{S}^\dagger\cdot\vec{\epsilon}|^2
 \equiv\frac{1}{2}\left\{\frac{5}{9}\vec{k}^2-\frac{1}{3}\vec{k}^2\cos^2\theta_1\right\}
\end{align}
where $\theta_1$ is the angle between the photon and the $\pi^0$.
The variables $\tilde{p}_2$, $\tilde{p}_3$, defined in the $2+3$ rest frame, are
conveniently boosted to the $\gamma p$ rest frame in order to evaluate
the invariant masses entering the evaluation of $T$.
Summing the tree level amplitude and the triangle diagram, $T$ in Eq.~(\ref{eq_13_1}) is given by
\begin{align}
 -i\,T=&(a+b)\,\vec{S}\cdot\vec{k} \; \vec{S}^\dagger\cdot\vec{\epsilon},
\end{align}
where
\begin{align}
 a = &\, C \; \frac{1}{M_{\rm inv}(12)-M_\Delta+i\Gamma_\Delta/2} \label{eq:facA}\\
 b = &\, C \; 2M_N \,2M_\Delta \, t_T \, t_{\eta p,\eta p}\\
 C = &\, g_{\Delta^*,\eta \Delta} \, g_{\Delta^*,\gamma N}
 \, \frac{f_{\pi N \Delta}}{m_\pi} \sqrt{\frac{2}{3}}
 \, \frac{1}{\sqrt{s} -M_{\Delta^*} +i\,\Gamma_{\Delta^*}/2}.
\end{align}
Thus,
\begin{align}
 \overline{\sum}\sum|T|^2=|a+b|^2\, \frac{1}{2}\,
 \left\{\frac{5}{9}\vec{k}^2 -\frac{1}{3}\vec{k}^2\cos^2\theta_1\right\}.\label{eq_14_1}
\end{align}
In Eq.~(\ref{eq:facA}) we also employ the energy-dependent width of Eq.~(\ref{eq:EdepWidth}), but now as a function of the invariant mass $M_{\rm inv}(12)$.

We shall also see the contribution of the $\Delta^* \to \eta \Delta$ alone, just taking $|a|^2$ in
Eq.~(\ref{eq_14_1}), and of $\pi N(1535)$, taking $|b|^2$ in Eq.~(\ref{eq_14_1}), instead of
$|a+b|^2$, which corresponds to the coherent sum of the two processes.

\begin{figure}[H]
 \centering
 \includegraphics[width=0.4\textwidth]{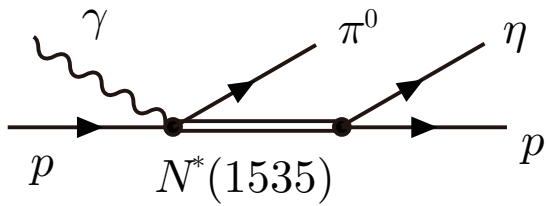}
 \caption{Standard mechanism for $\gamma p \to \pi N(1535)$ production.}
 \label{fig_3}
\end{figure}

The singularity that comes out from $t_T$ in Eq.~(\ref{eq_12_1}) leads to a peculiar energy dependence of the cross section for $\pi N(1535)$ production. In order to show it, we also evaluate the $\gamma p \to \pi N(1535)$ cross section with a standard mechanism that does not involve the triangle singularity, and which we show in Fig.~\ref{fig_3}.

The transition from angular momentum $1^- + 1/2^+ \to 0^- + 1/2^-$ requires $P$-wave, $L=1$, to restore the parity. Two structures are possible,
\begin{align}
 \vec{\epsilon}\cdot\vec{k},\ (\vec{\sigma}\times\vec{k})\cdot\vec{\epsilon}\,.
\end{align}
Both of them, after squaring and summing over polarizations, taking into account the transversality of the photons, Eq.~(\ref{eq_13_2}), lead to a combination
\begin{align}
 \overline{\sum}\sum|t|^2 \propto c\,\vec{k}^2
 +d\,\vec{k}^2\cos^2\theta_1
 \sim(c +\frac{1}{3}d\,)\vec{k}^2
\end{align}
where in the last step we have substituted $\cos^2\theta_1$ by $1/3$ as
it would come by integration over the phase space. Then, the cross section from this mechanism can be obtained from Eq.~(\ref{eq_13_1}) substituting $|T|^2$ by a constant times $\vec{k}^2| t_{\eta p,\eta p}|^2$
\begin{align}
 |T|^2\to D\,\vec{k}^2|t_{\eta p,\eta p}|^2.
\end{align}

\section{Results}

\begin{figure}[h!]
 \centering
  \includegraphics[width=0.5\textwidth]{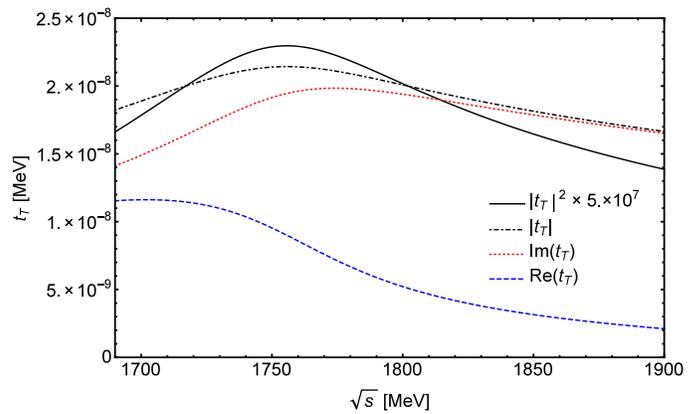}
 \caption{$|t_T|^2$, $Re(t_T)$, $Im(t_T)$ and $|t_T|$ as a function of the $\gamma p$
 energy, $\sqrt{s}$. The mass of the $\eta p$ system is taken at the $N(1535)$ mass of 1543 MeV determined in \cite{Inoue:2001ip}.}
 \label{fig_4}
\end{figure}

In Fig.~\ref{fig_4}, we show the result for the amplitude $t_T$ of Eq.~(\ref{eq_12_1}). We can see that $Re(t_T)$ and $Im(t_T)$ have the Breit-Wigner shape like $-BW \equiv -(\sqrt{s} -m_R +i\,\Gamma_R/2)^{-1}$. However, the real part does not go through zero, so the shape resembles $D-BW(s)$, with $D$ a constant real background. We can see in $|t_T|^2$ a clear peak around $\sqrt{s}=1770$ MeV, as anticipated by the simple application of the rule found in Ref. \cite{Bayar:2016ftu}.

\begin{figure*}[t!]
 \centering
 \includegraphics[width=\textwidth]{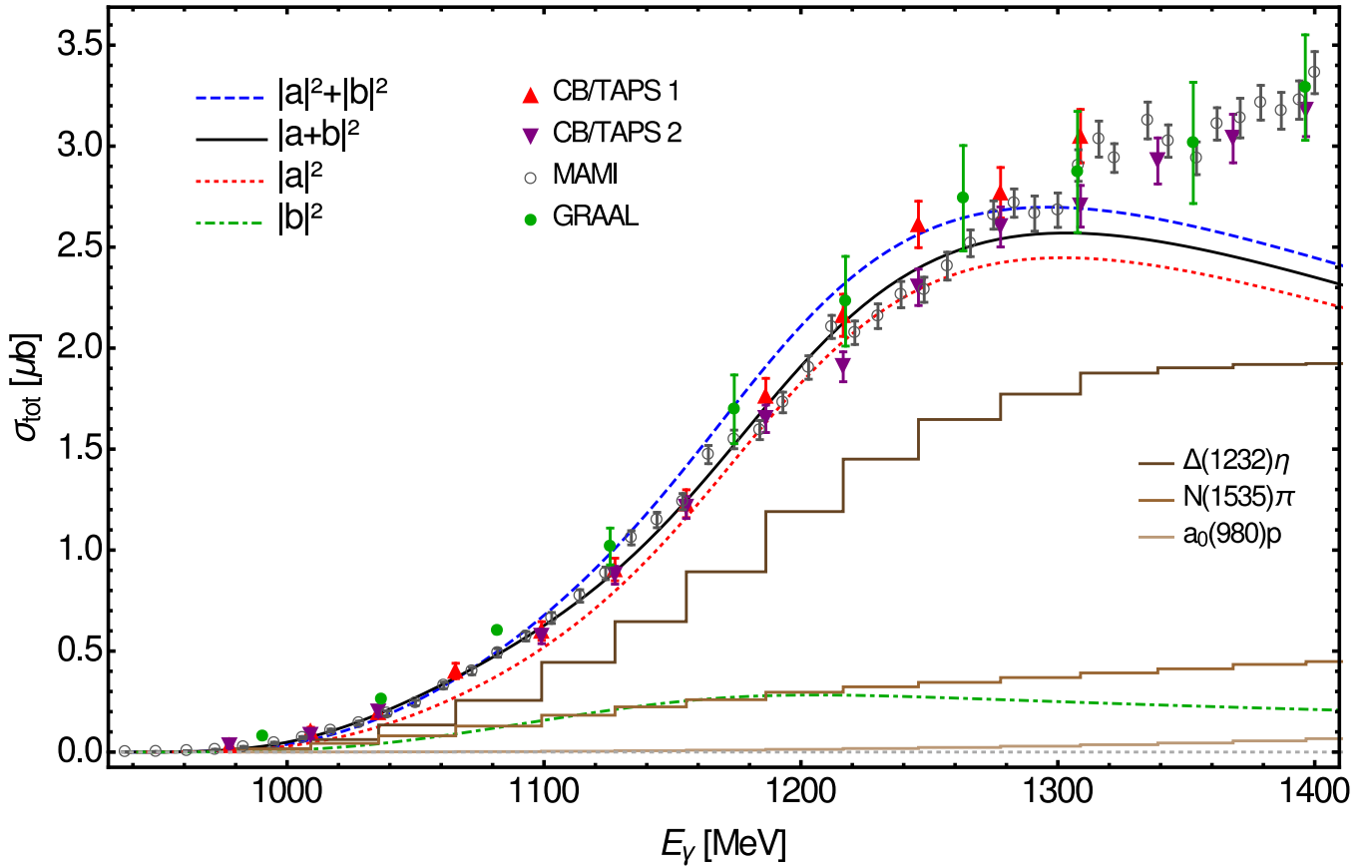}
 \caption{Cross section for $\gamma p \to \pi^0 \eta p$.}
 \label{fig_5}
\end{figure*}

In Fig.~\ref{fig_5} we show the results for the cross section. The $t_{\eta p,\eta p}$ amplitude is taken from Ref. \cite{Inoue:2001ip} as a Breit-Wigner amplitude
\begin{align}
 t_{\eta p,\eta p} = \frac{g_{N^*,\eta p}^2}{M_{\rm inv}(\eta p) -M_{N^*} +i\,\Gamma_{N^*}/2},
\end{align}
and we take the values for $M_{N^*}$, $\Gamma_{N^*}$ and $g_{N^*,\eta p}$ from that work, which provides a fair reproduction of the scattering data,
\begin{align}
 g_{N^*,\eta p} = & \;1.77\,,\notag\\
 M_{N^*} = & \; 1543\ {\rm MeV},\notag\\
 \Gamma_{N^*} = & \; 92\ {\rm MeV}.\notag
\end{align}
The width seems a bit smaller compared to the PDG average $150$ MeV, but
in agreement with BES data $95\pm 25$ MeV \cite{Bai:2001ua} and not far from
the most recent determination of $120\pm 10$ MeV in Ref. \cite{Sokhoyan:2015fra}.

\begin{figure}[h!]
 \centering
 \includegraphics[width=0.5\textwidth]{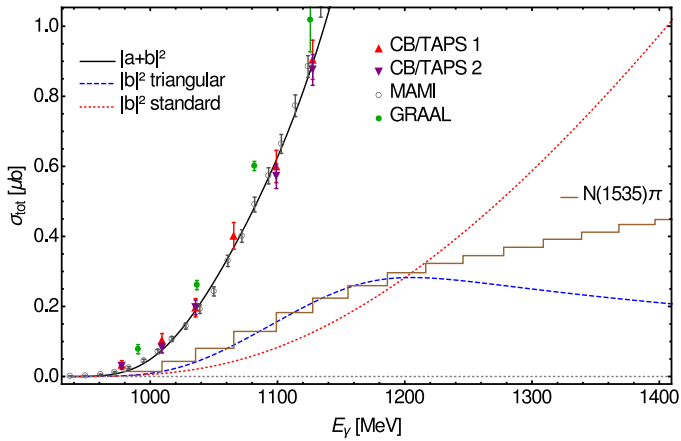}
 \caption{Cross section for the $\gamma p \to \pi N(1535)$ with the triangle mechanism and the standard mechanism of Fig.~\ref{fig_3}.}
 \label{fig_6}
\end{figure}

As shown in the former section, there are uncertainties in the mass, width and radiative decay of the $\Delta(1700)$. Playing with these uncertainties, one obtains a band of allowed cross sections from the dominant $\gamma p \to \Delta(1700) \to \eta \Delta \to \eta \pi^0 p$ mechanism, which is shown in Fig.~2 of Ref. \cite{Ajaka:2008zz}. Since we want to see the relative weight of the $\pi N(1535)$ production versus $\eta \Delta(1232)$, we fine tune the values of $M_{\Delta^*}$, $\Gamma_{\Delta^*}$ and $g_{\Delta^*,\gamma p}$ to get a fair agreement with the data for low energies. We find that the values that better fit the curve $|a+b|^2$ to the MAMI data \cite{Kashevarov:2009ww} up to 1300 MeV are $M_{\Delta^*}=1663.6$ MeV, $\Gamma_{\Delta^*} = 114.1$ MeV and $g_{\Delta^*,\gamma p}=0.142$, which corresponds to the branching fraction of 0.60\%.

 We obtain a fair reproduction of the cross section up to about $p_\gamma=1300$ MeV. From there on one would be relatively away from the $\Delta(1700)$ mass and other mechanisms discussed in Ref. \cite{Gutz:2014wit} should come into play. The important finding concerning the triangle singularity is that we obtain a $\pi N(1535)$ contribution in fair agreement with the experimental determination. One should not overstate the agreement, since the methods to obtain it in Ref. \cite{Gutz:2014wit} and here are different. In any case, the approximate agreement is welcome.

It is instructive to see that the cross section for $\pi N(1535)$ production is much wider than one could anticipate from the shape of $|t_T|^2$ in Fig.~\ref{fig_4}. This is because the $\pi^0 \eta p$ production amplitude has an extra factor $|\vec{k}|$ plus phase space factors and the weight of the $\Delta(1700)$ propagator.

In order to see the differences between the approaches followed here and in Ref. \cite{Gutz:2014wit} we show in Fig.~\ref{fig_5} the contributions of Eq.~(\ref{eq_13_1}) and Eq.~(\ref{eq_14_1}), taking $|a|^2$ (only $\eta \Delta$), $|b|^2$ (only $\pi N(1535)$) and $|a+b|^2$ in the equations. We can see that there is actually not much interference between the two amplitudes. Actually we find a small destructive interference, but this can become slightly constructive with small change of the parameters. The message is small interference, which happens in spite of the same spin structure of the two amplitudes as we have shown in Eqs.~(\ref{eq_10_1}) and (\ref{eq_11_1}). The reason is that the $N(1535)$ structure provided by the $t_{\eta p,\eta p}$ amplitude is multiplied by $t_T$, which as seen in Fig.~\ref{fig_4} has by itself a rich complex structure. This can explain the differences with the analysis of Ref. \cite{Gutz:2014wit}, where a more constructive interference between the two mechanisms occurs, as can be seen in Fig.~19 of that paper. In Ref. \cite{Gutz:2014wit} a partial wave analysis is done using a $K$-matrix approach in which the $t$-matrix is given $A_{ab}=K_{ac}(1-\rho K)^{-1}_{cb}$, where $\rho$ is a diagonal matrix that takes into account phase space of the intermediate states, and the kernel $K_{ac}$ is
written as background plus a sum of Breit-Wigner amplitudes, $\displaystyle K_{ab}=\sum_\alpha\frac{g_a^\alpha \, g_b^\alpha}{M_\alpha^2 -s} +f_{ab}$. It is clear that in this analysis there is no room for the analytically rich multiplicative structure of the triangle singular mechanism that we have studied here.

As commented at the end of the former section, we would like to show the effect of having the $\pi N(1535)$ production from the triangle mechanism. For that purpose, we compare in Fig.~\ref{fig_6} the results of our approach with the results that we would obtain using the mechanism of Fig.~\ref{fig_3}, for a standard production mechanism.

To facilitate the comparison we have normalized the cross sections at $E_\gamma=1200$ MeV. We can see that the shape of the cross section with the standard mechanism is quite different, and produces a cross section that keeps rising and has a concave shape. The mechanism that we have produces a different structure and gradually decreases around $E_\gamma=1300$ MeV, producing a better agreement with the experimental extraction in the range up to about $E_\gamma=1400$ MeV.

In Ref. \cite{Gutz:2014wit} the data are given up to $E_\gamma=2500$ MeV, but these are energies too big to contrast our model where only the $\Delta(1700)$ resonance excitation is included, together with a $\Delta(1700)$ induced triangle singularity to account for the $\pi N(1535)$ production. Other resonances and other mechanisms are at play at these energies \cite{Gutz:2014wit}: the $\pi N(1535)$ channel at higher energies would also receive contribution from another triangle singularity involving $\Sigma^{*0} K^+$ in the intermediate state with $\Sigma^{*0} \to \pi^0 \Lambda$ and $K^+ \Lambda$ fusing to give the $N(1535)$. Using the method of Ref. \cite{Bayar:2016ftu} the singularity peaks around $E_\gamma=1410$ MeV, about 200 MeV higher than the one we studied here. We have also evaluated the contribution of this singularity using the couplings from Refs. \cite{Sarkar:2004jh,Inoue:2001ip} and we find also a sizeable contribution above $E_\gamma=1400$ MeV, but for the purpose of the present work, its contribution is very small compared to the one we have calculated up to $E_\gamma=1300$ MeV, where with the limited information used here we give a fair description of the experimental data.

\section{Conclusions}

We have evaluated the $\gamma p \to \eta \pi^0 p$ cross section at low energies, up to about $500$ MeV above threshold for $E_\gamma$, taking into account two mechanisms: the $\gamma p \to \Delta(1700) \to \eta \Delta(1232) \to \eta \pi^0 p$ and the $\gamma p \to \Delta(1700) \to \pi N(1535)$. The first mechanism is the one shown to be dominant in the productions of Ref. \cite{Doring:2005bx} and subsequent papers. The second one is new and involves a triangle singularity in which $\Delta(1700) \to \eta \Delta(1232)$, $\Delta(1232) \to \pi^0 p$ and then $\eta p$ fuse to produce the $N(1535)$. The latter mechanism gave rise to a peak (broadened by the effect of the $\Delta(1232)$ width) around $E_\gamma=1220$ MeV ($\sqrt{s}=1782$ MeV). We have shown that this latter mechanism, which we can evaluate with elements borrowed from the properties of the $\Delta(1700)$ and $N(1535)$ as being dynamically generated resonances, gives rise to a contribution to the cross section in fair agreement with the experimental determination. We showed that the shape produced by the triangle mechanism is quite different from the one we would have assuming a standard $P$-wave $\pi N(1535)$ production mechanism, and the experimental determination is in better agreement with the triangle mechanism. We also showed that there is some discrepancy in the interference pattern between the two mechanisms with respect to the one obtained in Ref. \cite{Gutz:2014wit}, but we argued that this was a consequence of the fact that the analysis of Ref. \cite{Gutz:2014wit} does not include explicitly a triangle singularity in the approach. The triangle amplitude created by itself a kind of a resonance structure which multiplies (not sums) an amplitude like the one assumed in Ref. \cite{Kashevarov:2009ww}. As a consequence of this factor we find very small interference between the two mechanisms, while a more constructive interference is seen in the analysis of  Ref. \cite{Gutz:2014wit}. This also means that the amount of $\gamma p \to \eta \Delta(1232)$ in the cross section is somewhat bigger in our approach at energies around $E_\gamma=1100-1400$ MeV.

The exercise done here has also repercussion in other reactions. It has shown that in cases like the present one, where there is an unavoidable triangular singularity, the standard partial wave analysis
should be extended to accommodate such a structure. It is not clear a priori that a triangle singularity is going to have a relevance in a given reaction, but, given the simplicity of the rule developed in Ref. \cite{Bayar:2016ftu} to find out whether a singularity appears within a certain mechanism, it would be wise to make a general survey of a given reaction to see if such mechanisms can develop. One could easily derive the structure for this singularity, which up to a global factor only depends on the intermediate states of the triangle diagram, and use it as a multiplicative factor on top of the standard amplitude of present partial wave analysis. Analyses of data along these lines should be welcome in the future.

\section*{Acknowledgments}

We would like to thank E. Klempt and V. A. Nikonov for providing us the detailed experimental data and A. Ramos for pointing us the possible role of the singularity with $\Sigma^*$, $K$, $\Lambda$ intermediate states.

V. R. D. wishes to acknowledge the support from the Programa Santiago Grisolia of Generalitat Valenciana (Exp. GRISOLIA/2015/005).

S. S. wishes to acknowledge the support from the program Prometeo of the Generalitat Valenciana.

This work is also partly supported by the Spanish Ministerio de Economia y Competitividad and European FEDER funds under the contract number FIS2014-57026-REDT, FIS2014-51948-C2-1-P, and FIS2014-51948-C2-2-P, and the Generalitat Valenciana in the program Prometeo II-2014/068.

\clearpage

\bibliographystyle{plain}

\end{document}